# When the Quantum Energy Teleportation is Observable?

H. Razmi[(1)], A. MohammadKazemi[(2)]

Department of Physics, University of Qom, 3716146611, Qom, I. R. Iran.

(1) razmi@qom.ac.ir & razmiha@hotmail.com

(2) a.m.kazemi@gmail.com

**Abstract**

We want to investigate when the quantum energy teleportation is observable. The condition for observability is based on the minimum time value determined by the fundamental energy-time uncertainty relation.

**Introduction**

Quantum information theory is one of the products of a sum of curious works on the foundations of quantum physics from one side and some modern applications of it from the other side. Among a number of interesting techniques and protocols in this theory, Quantum Teleportation (QT) has been proposed to transfer quantum information between two distant points. QT is a modern and deep phenomenon around which scientists and engineers have been working for less than three decades. For the first time, the idea of QT (the standard protocol) was proposed in 1993 and experimentally confirmed in a short time [1-2]. Many theoretical and experimental research works have been already done on QT and it is rapidly in development now. One of the corresponding and exciting projects that has been proposed by the Japanese physicist Masahiro Hotta is the Quantum Energy Teleportation (QET) [3]. M. Hotta has already proposed a number of protocols for QET in different fields of physics [4-10]. In this paper, by considering one of the basic protocols of QET and another apparently important one, it is shown that the value of the teleported energy in the needed time for doing it, is at even smaller than the limit of the value constrained due to the uncertainty (indeterminacy) energy-time relation and thus it is unobservable (i.e. it cannot be really verified)!

**Quantum Energy Teleportation**

Let consider one of the basic models [11] in which the main idea behind all other protocols is seen. In summary, it is assumed that teleportation from site *A* to site *B* happens in a time much shorter than the usual energy transportation time. Considering an entangled system with two qubits *A* and *B*, the total Hamiltonian of this system is

$$H_{tot} = H_A + H_B + V, \qquad (1)$$

where

$$H_A = h\sigma_B^z + \frac{h^2}{\sqrt{h^2 + k^2}}, \qquad (2)$$

$$H_B = h\sigma_B^z + \frac{h^2}{\sqrt{h^2 + k^2}}, \qquad (3)$$

$$V = 2k\sigma_A^x \sigma_B^x + \frac{2k^2}{\sqrt{h^2 + k^2}}, \qquad (4)$$

and *h* and *k* are positive constants with energy dimensions.

The entangled ground state, $|g\rangle$, is given by

$$|g\rangle = \frac{1}{\sqrt{2}}\sqrt{1 - \frac{h}{\sqrt{h^2+k^2}}}|+\rangle_A|+\rangle_B - \frac{1}{\sqrt{2}}\sqrt{1 + \frac{h}{\sqrt{h^2+k^2}}}|-\rangle_A|-\rangle_B. \qquad (5)$$

A projective measurement is performed on this ground state,

$$P_A(\mu) = \frac{1}{2}(1 + (-1)^\mu \sigma_A^x). \qquad (6)$$

This measurement infuses energy, $E_A$, into site A

$$E_A = \frac{h^2}{\sqrt{h^2+k^2}}. \qquad (7)$$

The infused energy on site A diffuses into site B as

$$\langle H_B(t) \rangle = \frac{h^2}{2\sqrt{h^2+k^2}}[1 - \cos(4kt)]. \qquad (8)$$

Therefore, energy can be extracted from B after a diffusion time scale of $1/k$ (usual transportation time). Soon after the information about the measurement result is sent, via a classical channel, to site B, another measurement is performed on this site on the basis of this information. Now positive amount of energy

$$E_B = \frac{h^2 + 2k^2}{\sqrt{h^2+k^2}}\left[\sqrt{1 + \frac{h^2 k^2}{(h^2+2k^2)^2}} - 1\right]. \qquad (9)$$

can be extracted on this site. Teleportation occurs!

**Is this Quantum Energy Teleportation experimentally observable?**

In what follows and according to the energy-time uncertainty relation, it is shown that the above-mentioned QET protocol will not be observable and cannot be verified experimentally.

Since $h$ and $k$ are constants we can assume,

$$h = \alpha k. \qquad (10)$$

This leads to,

$$E_B = f(\alpha)k, \qquad (11)$$

where

$$f(\alpha) = \frac{\alpha^2 + 2}{\sqrt{\alpha^2+1}}\left[\sqrt{1 + \frac{\alpha^2}{(\alpha^2+2)^2}} - 1\right]. \qquad (12)$$

It can be easily shown that $f(\alpha)$ is less than 1 (one can plot $f(\alpha)$ versus $\alpha$ and find its maximum value about 0.13 ($Max(f(\alpha)) \sim 0.13$) as in figure 1).

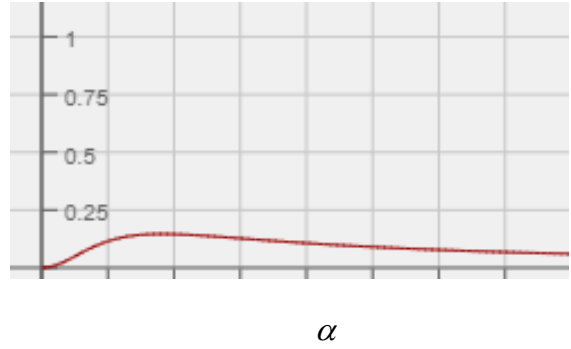

*Figure 1: Plot of $f(\alpha)$ relative to $\alpha$.*

So,

$$E_B \leq 0.13k \quad (13).$$

Hotta has correctly considered this fact that the teleportation time should be much shorter than the usual transportation time $1/k$ ($t_{eleportation} << 1/k$). Using this fact and the energy-time uncertainty relation $E_B t_{teleportaion} \geq 1$***, it is found:

$$1 << 0.13 \quad !? \quad (14)$$

This is a contradictory result; the above-proposed QET protocol isn't observable.

**Quantum energy teleportation with trapped ions**

Now, let investigate another QET proposal corresponding to trapped ions [5]. In this proposal, the teleported energy is:

$$E_{out} = \gamma_N E_{in} \exp(-\zeta_N \frac{E_{in}}{\upsilon}) \sin^2(2\varphi). \quad (15)$$

Knowing that with increasing the number of cold ions N, the coefficient $\gamma_N$ decreases exponentially ($\propto e^{-1/N}$), it can be simply found that the maximum value of the output

(teleported) energy is realized when the input energy is of the order of the energy of a phonon in the ion crystal ($h\upsilon$); this means:

$$E_{out}^{max.} < \upsilon \exp(-\zeta_N) \Rightarrow E_{teleportaion} << \upsilon. \quad (16)$$

The energy carrier in this protocol is the phonon energy; thus, the usual energy transmission (transportation) time is of the order of the inverse of the frequency which may be considered as an upper bound for the QET time:

$$t_{teleprtation} < \frac{1}{\upsilon}. \quad (17)$$

Considering (16) and (17), it is found that

$$E_{teleportaion} t_{teleportaion} << 1 \quad ?! \quad (18)$$

which is in contradiction to the energy-time uncertainty relation $E_{teleportaion} t_{teleportaion} \geq 1$*** that should be verified as the fundamental condition for the observability of the QET protocol under consideration.

**Conclusion**

Although it has been claimed that the theoretically proposed QET idea can be experimentally verified [10], in this paper, two considerable models of QET including a basic protocol and another one with trapped ions have been investigated and it has been shown that the teleported energy value isn't in the observationally allowed region specified by the fundamental energy-time uncertainty relations. Not only about the already proposed QET protocols, but also for any other model, the restrictions applied by the fundamental postulates of quantum theory particularly those are based on the observability criterions by the uncertainty relations must be considered.

\*\*\*

*This is a simply well-known fact that when one works with a quantity as A, its uncertainty should be less than it $\langle A \rangle \geq \Delta A$; otherwise, the resulting value isn't experimentally acceptable/reliable. This is why the inequality $E_B t_{teleportaion} \geq 1$ implicitly is based on $\Delta E_B \Delta t_{teleportaion} \geq 1$ because $E_B \geq \Delta E_B$ and $t_{teleportaion} \geq \Delta t_{teleportaion}$. The relations (13) and (16) ($E_B \leq 0.13k$ and $E_{out}^{max.} < \upsilon \exp(-\zeta_N) \Rightarrow E_{teleportaion} << \upsilon$) reason more strongly on this fact that $\Delta E_B \leq 0.13k$ and $\Delta E_{teleportaion} << \upsilon$; similarly about the teleportation time.*